\documentclass[twocolumn,showpacs,preprintnumbers,amsmath,amssymb]{revtex4}
\usepackage{indentfirst}
\usepackage{hhline}
\usepackage{wrapfig, float}
\usepackage{array,longtable,tabularx}
\usepackage{caption}
\usepackage{graphicx}
\usepackage{dcolumn}
\usepackage{bm}
\usepackage{subfigure}
\usepackage{color}
\usepackage{hyperref}
\hypersetup{linkcolor=blue, citecolor=blue, urlcolor=blue, colorlinks=true}

\begin{document}


\preprint{1412.0497} 

\title {Suppression of Superconductivity in Lu$_x$Zr$_{1-x}$B$_{12}$ : Evidence of Static Magnetic Moments Induced by Non-Magnetic Impurities}

\author{N.E.Sluchanko$^{1}$}\email{nes@lt.gpi.ru}\author{A.N.Azarevich$^{1}$} \author{M.A.Anisimov$^{1}$}
\author{A.V.Bogach$^{1}$} \author{S.Yu.Gavrilkin$^{2}$} \author{V.V.Glushkov$^{1,3}$}
\author{S.V.Demishev$^{1,3}$}\author{A.L.Khoroshilov$^{3}$}\author{A.V.Dukhnenko$^{4}$}\author{K.V.Mitsen$^{2}$}
\author{N.Yu.Shitsevalova$^{4}$}\author{V.B.Filippov$^{4}$} \author{V.V.Voronov$^{1}$}
\author{K.Flachbart$^{5}$}

\affiliation{$\phantom{x}^1$ $-$ A.M.Prokhorov General Physics
Institute of RAS, 38 Vavilov Street,119991 Moscow Russia}

\affiliation{$\phantom{x}^2$ $-$ P.N.Lebedev Physical Institute of RAS, Moscow, 119991 Russia}

 \affiliation{$\phantom{x}^3$--Moscow Institute of Physics and Technology, 141700 Dolgoprudnyi, Russia}

\affiliation{$\phantom{x}^4$ $-$ Institute for Problems of Materials
Science of National Academy of Sciences of Ukraine, \\ 3 Krzhizhanovskogo Street, 03680 Kiev, Ukraine}

\affiliation{$\phantom{x}^5$ $-$ Institute of Experimental Physics of
SAS, 47 Watsonova Street, SK-04001 Kosice, Slovak Republic}

\date{\today}
\begin{abstract}
Based on low temperature resistivity,
heat capacity and magnetization investigations
we show that the unusually strong suppression
of superconductivity in Lu$_x$Zr$_{1-x}$B$_{12}$
BSC-type superconductors in the range $x$$<$0.08 is caused by the emergence
of static spin polarization in the vicinity of
non-magnetic lutetium impurities. The analysis of
received results points to a formation of static
magnetic moments with $\mu_{eff}$$\approx$$3\mu_B$ per Lu-ion. The
size of these spin polarized nanodomains was estimated to be about 5 ${\AA}$.
\end{abstract}

\maketitle

\textbf{1.} The discovery of superconductivity at $T$$_C$$\approx$39 K in
MgB$_2$ \cite{1} stimulated a significant interest into the studies
of a wide class of the alkaline-earth, rare-earth and transition-metal
borides. Among them, in the family of higher borides RB$_{12}$, zirconium
dodecaboride (ZrB$_{12}$) is a BCS superconductor with the highest $T$$_C$$\approx$6 K \cite{2,3}.
An intriguing detail established for ZrB$_{12}$ is the formation of Cooper pairs
through quasi-local vibrations involving Zr$^{4+}$ ions located within truncated
$B$$_{24}$ octahedrons in the UB$_{12}$-type \textit{fcc} crystal structure \cite{2}-\cite{5}. In studies of
the Einstein phonon mediated superconductivity in ZrB$_{12}$ authors of \cite{2}-\cite{9}
argue that \textit{s}-wave pairing is characteristic for this compound, and that in
this case the Ginzburg-Landau parameter $\kappa$ is located in the nearest vicinity
of the threshold value $\kappa_C$$=$2$^{-1/2}$. Moreover, a crossover from type-I to
type-II$/$1 superconductivity with temperature lowering was deduced in \cite{3}
from heat capacity and magnetization measurements. In contrast, in \cite{10}
the superconductivity in ZrB$_{12}$ was interpreted in terms of \textit{d}-wave pairing
and a two-gap type-II regime was identified with parameters $\kappa_p$$=$3.8 and
$\kappa_d$$=$5.8. Additionally, a large size pseudo-gap ($\Delta$$\sim$7.3 meV) has been
detected employing high resolution photoemission spectroscopy in ZrB$_{12}$
above $T$$_C$, and proximity to the quantum fluctuation regime was predicted
from \textit{ab initio} band structure calculations \cite{11}. Thus, also similarities
with cuprate high temperature superconductors (HTSC) may be supposed to
revive the interest into studies of this low temperature superconductor.

In case of non-magnetic impurity substitutions and their impact on
superconducting properties one can choose between various scenarios,
because for these defects the pair-breaking mechanism in various models
differs. According to Anderson's theorem \cite{12,13} and its extension to
non-\textit{s}-wave superconductivity (see e.g. \cite{14,15}) already a small amount
of non-magnetic impurities can dramatically suppress superconductivity
by pair-breaking in the case of anisotropic gap in a \textit{d}-wave superconductor.
Moreover, experiments on cuprates reveal \cite{15} that a spinless impurity
(Zn, Li etc.) introduced into a HTSC host produces in its vicinity a large
and spatially extended alternating magnetic polarization. On the basis of
NMR and $\mu$SR spectra it has been demonstrated that this impurity-induced
magnetization on the nearest neighbor Cu atoms in cuprates is associated
with a dynamic moment \cite{15}.

	In the case of RB$_{12}$ the replacement of non-magnetic ions of Zr by
Lu produces an about 15 times' reduction of superconducting transition
temperature ($T$$_C$$\approx$0.4 K for LuB$_{12}$ \cite{4,9,16}), and the origin of this
large $T$$_C$ suppression is not cleared up to now for these two compounds
with similar conduction bands and crystalline structures. Indeed,
inelastic neutron scattering studies of the phonon spectra in LuB$_{12}$
and ZrB$_{12}$ \cite{5} have detected noticeable, but not dramatic changes in
the position of almost dispersion-less quasi-local mode (15 meV and
17.5 meV, correspondingly), which was proposed to be responsible for
Cooper pairing. Only a moderate difference in the electronic density
of states of these two compounds is caused by filling of a wide enough
conduction band ($\sim$1.6$\div$2 eV) when Lu$^{3+}$ $-$ ion is changed to Zr$^{4+}$ in the RB$_{12}$
unit cell, resulting to an elevation by about 0.3-0.4 eV of the Fermi
level $E$$_F$ for ZrB$_{12}$ in comparison to LuB$_{12}$ \cite{11,17}.

\begin{figure}[htpb]
\begin{center}
\includegraphics[width = 7.3cm]{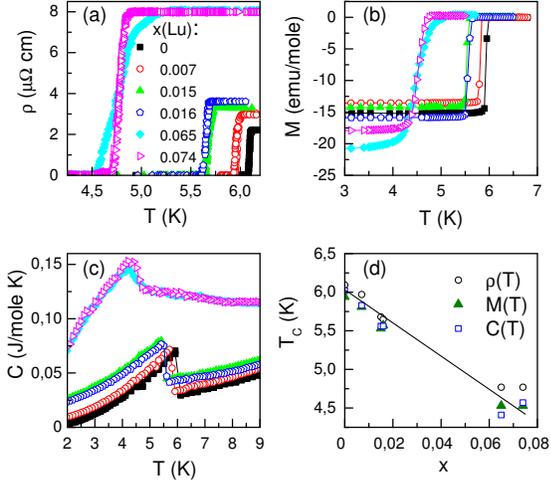}
   \caption{(Colour on-line) Temperature dependences of (a) resistivity, (b) magnetization at $H$$=$5 Oe and (c) specific heat in vicinity of \textit{T$_C$}, and (d) the suppression of superconductivity \textit{T$_C$(x)} in Lu$_x$Zr$_{1-x}$B$_{12}$.}\label{FigX1}
   \end{center}
\end{figure}

	Here we probed the evolution of superconducting transition
temperature $T$$_C$ and the normal state parameters for substitutional
solid solutions Lu$_x$Zr$_{1-x}$B$_{12}$ employing resistivity, heat capacity
and magnetization measurements. It will be shown that the non-magnetic
Lu impurity substitution (having a 4$f^{14}$ configuration) produces a
strong static spin polarization in the RB$_{12}$ matrix in the vicinity
of lutetium ions. Simultaneously with the emergence of static magnetic
moments with a value of about 3$\mu_B$ per Lu ion the received experimental
results exhibit a strong suppression of superconductivity in Lu$_x$Zr$_{1-x}$B$_{12}$.

\textbf{2.} Studies of resistivity, transverse magnetoresistance,
heat capacity and magnetization of high-quality single crystals of
Lu$_x$Zr$_{1-x}$B$_{12}$ solid solutions with $x$$<$0.08 were carried out at temperatures
in the range 1.8-300 K, in magnetic fields of up to 90 kOe (\textbf{H}$\|$$\langle$001$\rangle$).
A standard \textit{dc} four probe technique was applied for resistivity investigation
with the orientation of measuring current \textbf{I}$\|$$\langle$110$\rangle$.
Magnetization and heat capacity were measured using a PPMS-9 (Quantum Design).
The single crystals of Lu$_x$Zr$_{1-x}$B$_{12}$ were grown by vertical crucible-free inductive
floating zone melting in an inert gas atmosphere.
To verify the samples' quality and the Lu content X-rays Laue back-pattern (see Fig.S1 in the \hyperref[SI]{Supplementary Information}) and microanalysis techniques were applied.

\begin{figure}[htpb]
\begin{center}
\includegraphics[width = 8.4cm]{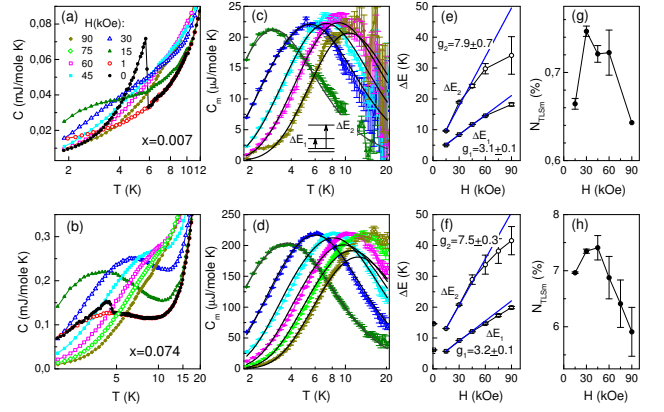}
   \caption{(Colour on-line) Temperature dependences of specific heat for Lu$_x$Zr$_{1-x}$B$_{12}$ with (a) $x$$=$0.007 and (b) $x$$=$0.074 in magnetic field up to 90 kOe. Panels (c) and (d) demonstrate the separated magnetic component $C$$_m$ of heat capacity for Lu content $x$$=$0.007 and $x$$=$0.074, correspondingly. The fitting of $C$$_m$\textit{(T,H$_0$)} by three-level Schottky relation (see scheme on panel (c)) is shown by solid lines. The splitting energies  $\Delta E_1$\textit{(H)}, $\Delta E_2$\textit{(H)} and concentrations $N$$_{TLSm}$\textit{(H)} are shown for Lu$_x$Zr$_{1-x}$B$_{12}$ with $x$$=$0.007 and 0.074 on panels (e),(f) and (g),(h), correspondingly.}\label{FigX2}
   \end{center}
\end{figure}

\textbf{3.} Temperature dependences of resistivity (fig.\hyperref[FigX1]{1a}),
magnetization (fig.\hyperref[FigX1]{1b}) and heat capacity (fig.\hyperref[FigX1]{1c}) show superconducting
phase transitions with $T$$_C$ in the range 4.5-6 K for Lu$_x$Zr$_{1-x}$B$_{12}$ ($x$$<$0.08)
solid solutions. The resistivity  \textit{$\rho$(T)} drop to zero below $T$$_C$ is accompanied
both with the appearance of Meissner state diamagnetic response on \textit{M(T)}
curves and the stepwise changes in the specific heat \textit{C(T)} (see fig.\hyperref[FigX1]{1} and
figs.\hyperref[FigX2]{2a-b}, $H$$=$0 curves). With the increase of lutetium content both the residual
resistivity and the normal state specific heat rise dramatically (by a factor of 4,
see fig.\hyperref[FigX1]{1a} and figs.\hyperref[FigX2]{2a} and \hyperref[FigX2]{2b}), and the later one demonstrates a combination of
superconducting step-like and Schottky-type anomaly (see e.g. fig.\hyperref[FigX2]{2b}, $H$$=$0 curve for
$x$$=$0.074). It is worth to note that the Schottky anomaly for Lu$_{0.074}$Zr$_{0.0926}$B$_{12}$
(fig.\hyperref[FigX2]{2b}, $H$$=$0) is very similar to that one observed previously for LuB$_{12}$ \cite{9,18,19},
and it may be interpreted in terms of formation of two-level systems (TLS) in the
disordered RB$_{12}$ matrix of this cage-glass compound. The \textit{T$_C$($x$)} dependence for
Lu$_x$Zr$_{1-x}$B$_{12}$ solid solutions is summarized in fig.\hyperref[FigX1]{1d}. It should be stressed
that the suppression of superconductivity by Lu substitution is unusually
strong ($\sim$0.21 K$/$at.$\%$ Lu), when supposing the doping of an
\textit{s}-wave superconductor by non-magnetic impurities.

In external magnetic field the amplitude of the low temperature Schottky
anomaly increases essentially and \textit{C(T)} maximum moves to higher temperatures
(for example, figs.\hyperref[FigX2]{2a}, \hyperref[FigX2]{2b} show the data for $x$$=$0.007 and 0.074, correspondingly).
Evidently, the Zeeman component is a magnetic contribution to heat capacity and
it may be separated from vibrational and electron heat capacity terms following
the approach developed in \cite{18,19}. The resulting $C$$_m$\textit{(T,H$_0$)} dependences are
presented in figs.\hyperref[FigX2]{2c},\hyperref[FigX2]{2d} for these two Lu contents together with their
approximation (see solid lines in figs.\hyperref[FigX2]{2c},\hyperref[FigX2]{2d}) based on a three-level Schottky relation (see eq.(S1) in \hyperref[SI]{supplementary materials})
\begin{figure}[htpb]
\begin{center}
\includegraphics[width = 7.7cm]{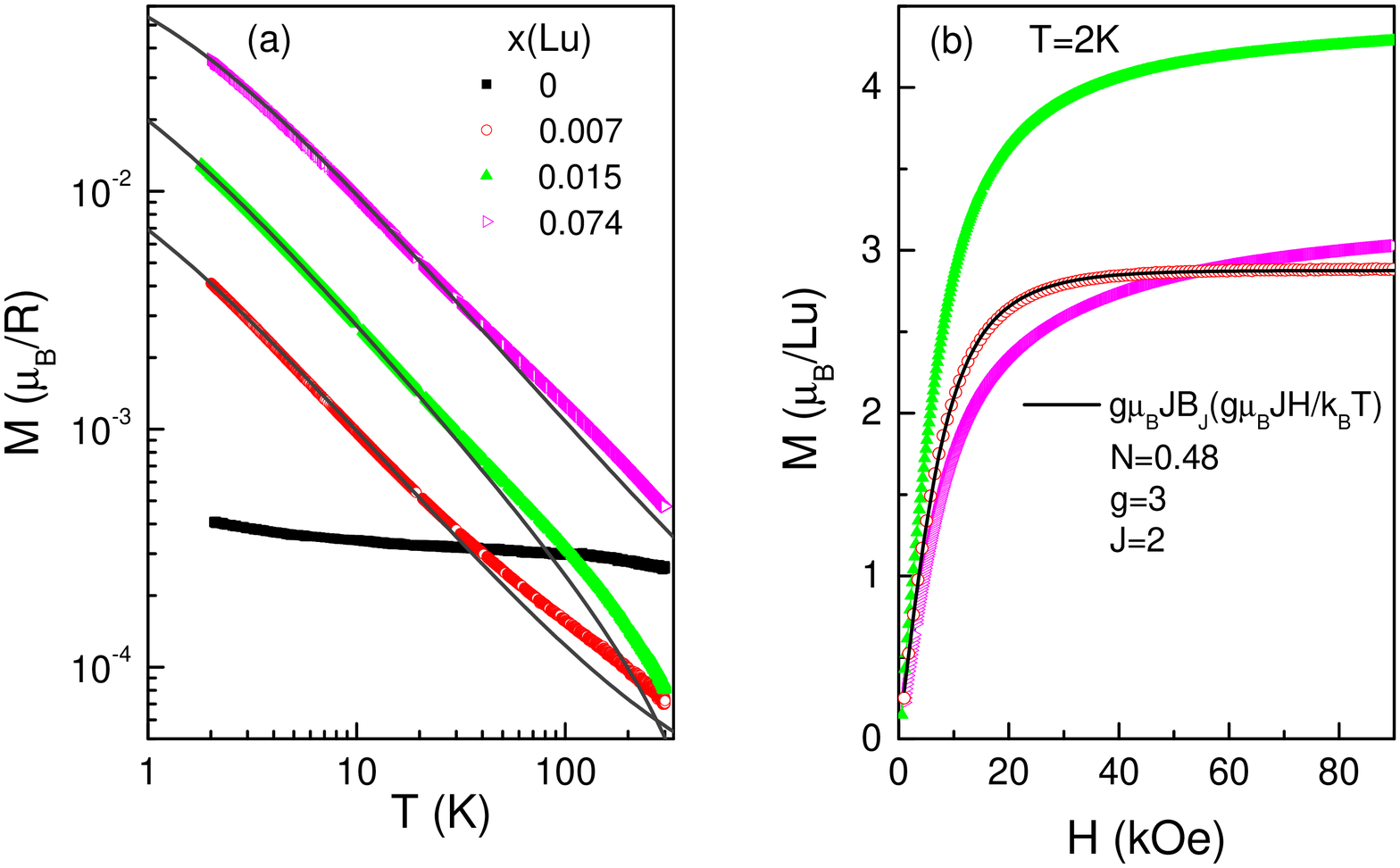}
   \caption{(Colour on-line) Dependences of magnetization vs. temperature at $H$$=$2 kOe (a) and vs. magnetic field at $T$$=$2 K (b) for Lu$_x$Zr$_{1-x}$B$_{12}$. Approximations by Curie-Weiss relation (\hyperref[Eq.2]{1}) (panel (a)) and by Brillouin dependence (\hyperref[Eq.3]{2}) (panel (b)) are shown by solid lines.}\label{FigX3}
   \end{center}
\end{figure}
(the scheme presented in insert to fig.\hyperref[FigX2]{2c}). Both splitting energies \textit{$\Delta$E$_1$(H)}, \textit{$\Delta$E$_2$(H)}
and the concentration of magnetic Schottky sites \textit{N$_{TLSm}$(H)} as deduced from the
approximation are shown on figs.\hyperref[FigX2]{2e}, \hyperref[FigX2]{2g} and figs.\hyperref[FigX2]{2f}, \hyperref[FigX2]{2h},  correspondingly.
The slope of straight lines on figs.\hyperref[FigX2]{2e} and \hyperref[FigX2]{2f} (in the range $H$$<$40 kOe)
allows to determine the g-factors from relation $\Delta E_i$$=$$\mu_B$g$_i$H$/k_B$ and obtain the values g$_1$$\sim$3.2 and g$_2$$\sim$7.8 for both studied crystals.
Moreover, the concentration of TLSm is found to be within experimental accuracy equal
to $x$ values in Lu$_x$Zr$_{1-x}$B$_{12}$. From these results it may be concluded that magnetic
sites are created by Lu substitution.

The magnetic response in the normal state of Lu$_x$Zr$_{1-x}$B$_{12}$ superconductors
was investigated by magnetization \textit{M(H,T)} studies. Figs.\hyperref[FigX3]{3a} and \hyperref[FigX3]{3b} demonstrate
the temperature dependences \textit{M(H$_0$,T)} recorded at $H$$_0$$=$2 kOe and the magnetization
vs. magnetic field curves \textit{M(H,T$_0$)} measured at $T$$_0$$=$2 K, correspondingly. It can
be seen from fig.\hyperref[FigX3]{3a}, that a small ($\sim$4$\times$10$^{-4}$ $\mu_B/$Zr) and about temperature
independent Pauli-like paramagnetic response, which is typical for ZrB$_{12}$,
changes into a Curie-Weiss-type magnetic signal originating from localized
magnetic moments induced by the Lu substitution. Fitting the temperature
dependences by Curie-Weiss relation

\begin{equation}\label{Eq.2}
M=\chi_0 H + \mu_{eff}^2 H/(k_B (T-\Theta))
\end{equation}
\begin{figure}[htpb]
\begin{center}
\includegraphics[width = 8.6cm]{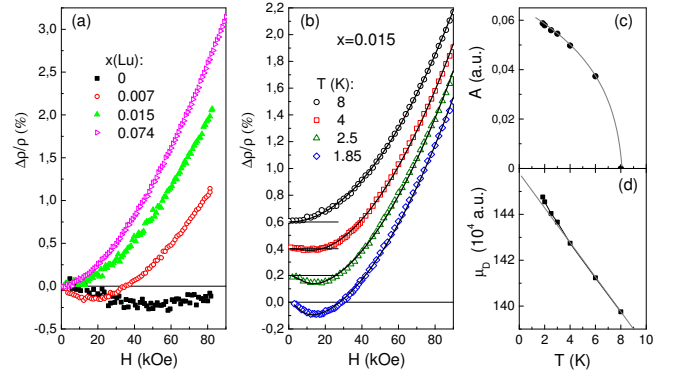}
   \caption{(Colour on-line) Magnetic field dependences of normal state magnetoresistance   $\Delta \rho / \rho$(H,T$_0$) (a) for Lu$_x$Zr$_{1-x}$B$_{12}$ with various Lu content at $T$$_0$$=$4.5 K, and (b) for $x$$=$0.015 at temperatures in the range 1.85-8 K. Approximations in the framework of (\hyperref[Eq.5]{eq.4}) are shown by solid lines in panel (b); curves are shifted by 0.2$\%$ for convenience. Panels (c) and (d) demonstrate temperature dependences of $A$ and $\mu_D$ parameters in (\hyperref[Eq.5]{eq.4}).}\label{FigX4}
   \end{center}
\end{figure} where $\chi_0$ - temperature independent susceptibility, $k$$_B$ - Boltzman constant and $\Theta$ - Curie-Weiss temperature (see solid lines in fig.\hyperref[FigX3]{3a}) allows to determine the Curie constant and subsequently
to estimate the effective magnetic moment $\mu_{eff}$$\sim$3.3-3.8$\mu_B$ per Lu ion (see also Tab.S1 in the \hyperref[SI]{Supplementary materials}).
These values correlate very well with the saturated moments $\mu_S$$\sim$3-4$\mu_B$
per Lu ion obtained from the analysis of magnetization vs. magnetic field dependences
(fig.\hyperref[FigX3]{3b}).

Moreover, in the regime of isolated magnetic impurity for $x$$=$0.007
the magnetization may be described with a good accuracy by the Brillouin-type dependence

\begin{equation}\label{Eq.3}
M=N_m g \mu_B J B_J(g \mu_B JH/k_B T)
\end{equation}
(where $N$$_m$$-$concentration of magnetic sites, $B$$_J$$-$Brillouin function, $\mu_B$$-$Bohr magneton and J$-$angular momentum) with $N$$_m$$\approx$0.48$x$, g$\approx$3 and J$\approx$2. As a result, it encourages us to suggest as a simplest
scenario that impurity ions of Lu are combined into dimmers with a magnetic
moment  $\mu_S$$=$gJ$=$6$\mu_B$ and with a size equal to about d(Lu-Lu)$\sim$5.3 ${\AA}$ $-$ the distance
between heavy ions in the RB$_{12}$ matrix, which can be considered as a formation of
spin-polarized nanodomains in Lu$_x$Zr$_{1-x}$B$_{12}$.

	Magnetoresistance $\Delta\rho/\rho$ experiments were also performed to characterize
the magnetic moments in these compounds with metallic conduction. According
to Yosida calculations carried out within the framework of \textit{s-d} exchange model,
an appearance of negative magnetoresistance (nMR) is expected in the regime
of charge carrier scattering on localized magnetic moments \cite{20}. Both field
and temperature dependences of nMR are controlled by local magnetization $M$$_{loc}$ through the relation

\begin{equation}\label{Eq.4}
    -\Delta\rho/\rho \sim M_{loc}^2.
\end{equation}
The nMR effect may be considered as an independent argument in favor
of local moments' formation in a metallic matrix. Fig.\hyperref[FigX4]{4} shows the results of
magnetoresisance measurements on Lu$_x$Zr$_{1-x}$B$_{12}$. In the normal state of ZrB$_{12}$
the main contribution to  $\Delta\rho/\rho$\textit{(H,T$_0$)} is positive and it can be described with a
good accuracy by the well-known relation  $\Delta\rho/\rho$$\sim$$\mu_D$$^2$$H$$^2$, where $\mu_D$ is the carriers' drift mobility.
Both the substitutional disorder and the cage-glass effect in Lu$_x$Zr$_{1-x}$B$_{12}$ decrease
dramatically both the mobility and amplitude of the positive component when Lu
concentration increases (fig.\hyperref[FigX4]{4a}). For $x$$\geq$0.015 the appearance of nMR contribution to $\Delta\rho/\rho$
becomes evident, and for $x$$\sim$0.07 the negative term in applied magnetic fields (fig.\hyperref[FigX4]{4a})
prevails. Fig.\hyperref[FigX4]{4b} presents a set of magnetic field dependences  $\Delta\rho/\rho$\textit{(H,T$_0$)} obtained for
$x$$=$0.015 at temperatures between 1.8 K and 8 K. In the range between 8 K and 4 K a
crossover from positive magnetoresistance to a combination of positive and
negative components is observed, and similar to the approach developed in \cite{21},
these data may be approximated very well by relation

\begin{equation}\label{Eq.5}
   \Delta\rho/\rho= -(A g \mu_B J B_J(g \mu_B JH/k_B T))^2+ \mu_D^2H^2.
\end{equation}
Fitting data by relation (\hyperref[Eq.5]{4}) (see solid lines in fig.\hyperref[FigX4]{4b}) allows to estimate the g-factor
g$\approx$3.5 and the angular momentum J$=$2. The obtained values are very close to those
deduced above from the analysis of magnetization results by relation (\hyperref[Eq.3]{2}).
Additionally, we have evaluated the behavior of coefficients \textit{A(T)} and
\textit{$\mu_D$(T)} in (\hyperref[Eq.5]{4}) (see figs.\hyperref[FigX4]{4c} and \hyperref[FigX4]{4d}, correspondingly). As can be seen from fig.\hyperref[FigX4]{4c},
the scattering of charge carriers on localized moments in vicinity of impurity
sites appears below 8 K and increases drastically with temperature lowering.
On the contrary, the  \textit{$\mu_D$(T)} dependence demonstrates only a moderate elevation
with temperature decrease (fig.\hyperref[FigX4]{4d}). Thus, these positive and negative
components of magnetoresistance become comparable at helium temperatures and
in magnetic fields below 40 kOe.

	When discussing the possible scenario of the formation of
spin-polarized nanodomains in this case, it is worth to note the
mechanism of spin-polaron formation proposed for the rare earth higher
borides with a cage-glass structure. It was suggested in \cite{22} that fast
quantum oscillations of heavy rare earth ions in a double-well potential
lead to spin polarization of \textit{5d}-conduction band states, and this effect
appears to be very sensitive to external magnetic field. Turning to the
analogy with non-magnetic impurity-induced \textit{dynamic} moments in HTSC cuprates,
it should be stressed here that the localized moments found in present study
of Lu$_x$Zr$_{1-x}$B$_{12}$ are \textit{static} and that a strong enhancement of spin polarization
(exhibiting a saturation above 40 kOe) is induced by external magnetic field.

	To summarize, we have found a formation of static magnetic moments with $\mu_{eff}$$\approx$3$\mu_B$ per Lu-ion in the vicinity of non-magnetic lutetium impurities in the Zr-rich matrix of Lu$_x$Zr$_{1-x}$B$_{12}$ dodecaborides at low temperatures. According to our opinion, the strong suppression of superconductivity in Lu$_x$Zr$_{1-x}$B$_{12}$ compounds should be attributed to pair-breaking arising in the vicinity of these nanosize magnetic domains.

We would like to thank A.V.Kuznetsov, G.E.Grechnev and S.Gabani for
helpful discussions. The study was supported by RFBR project No. 15$-$02$-$02553a,
Young Scientists Grant of the RF President No. MK$-$6427.2014.2. The measurements were carried out in Shared Facility Centre of P.N.Lebedev Physical Institute of RAS.


\newpage
\onecolumngrid
\section*{SUPPLEMENTARY INFORMATION}\label{SI}
\setcounter{figure}{0}
\renewcommand{\thefigure}{S\arabic{figure}}
1)
\begin{figure}[h]
 \begin{center}
\includegraphics[width = 5cm]{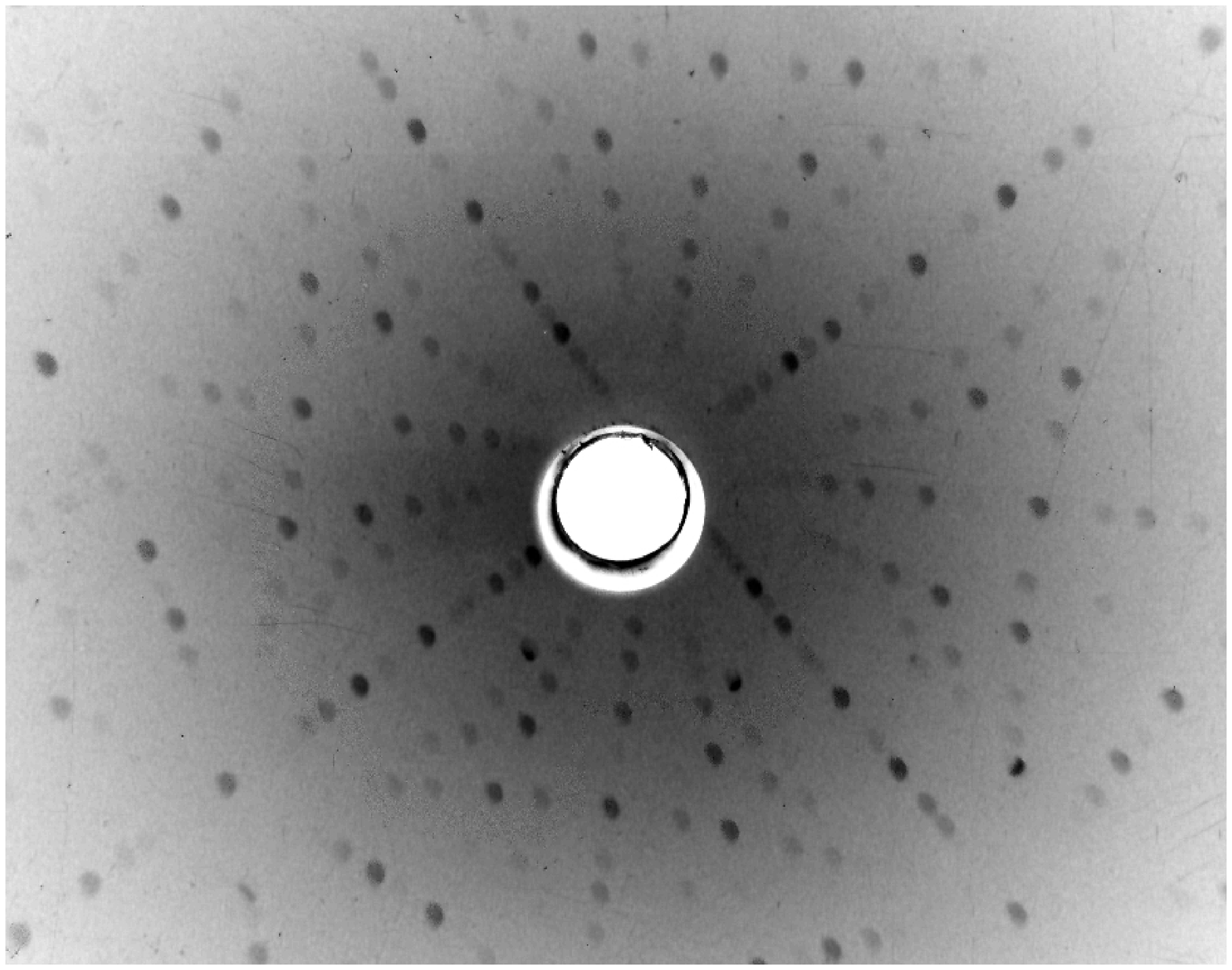}
    \end{center}
   \caption{X-ray Laue pattern from the lateral cross section of a raw single crystal Lu$_{0.074}$Zr$_{0.926}$B$_{12}$ grown with $[$100$]$ oriented seed. The growth direction deviation from $[$100$]$ is about 3 degree.}\label{FigXS1}
\end{figure}

\renewcommand{\theequation}{S1}
2) three-level Schottky relation:
\begin{equation}\label{Eq.S1}
   C_m = N \frac{R}{\beta^2} \frac{2 \Delta E_1^2 e^{-\beta \Delta E_1}+2 \Delta E_2^2 e^{-\beta \Delta E_2}+(\Delta E_1 - \Delta E_2)^2 e^{-\beta(\Delta E_1 + \Delta E_2)}}{(2+e^{-\beta \Delta E_1}+e^{-\beta \Delta E_2})^2}, \beta=1/k_B T
\end{equation}

3)

Tab.S1: The parameters of heat capacity and magnetization analysis performed for Lu$_x$Zr$_{1-x}$B$_{12}$ on figs.\hyperref[FigX2]{2} and \hyperref[FigX3]{3}.
\begin{figure}[h]
 \begin{center}
\includegraphics[width = 15cm]{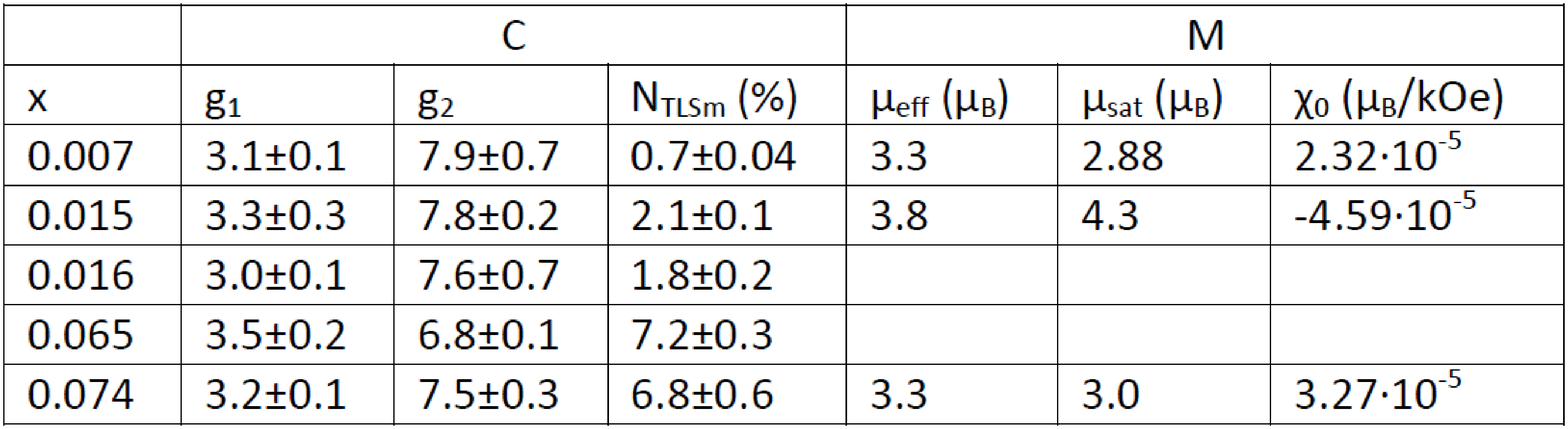}
    \end{center}
   \label{FigXS2}
\end{figure}

\end{document}